\documentclass[12pt]{article}

\usepackage{amsmath}
\DeclareMathOperator{\arccosh}{arccosh}
\usepackage{amsthm}
\usepackage{amssymb}
\usepackage{stmaryrd}
\usepackage{subfigure}
\usepackage{empheq}
\usepackage{xcolor}
\renewcommand{\theequation}{\arabic{equation}}
\newcommand{\EQ}{\begin{equation}}
\newcommand{\EN}{\end{equation}}

\newcommand{\bear}{\begin{eqnarray}}
\newcommand{\ear}{\end{eqnarray}}
\newcommand{\bt} { \begin{tabular} }
\newcommand{\et}{ \end{tabular} }
\newcommand{\bc} { \begin{center} }
\newcommand{\ec}{ \end{center} }

\newcommand{\btb} { \begin{table} }
\newcommand{\etb}{ \end{table} }

\begin{document}

\topmargin 0pt
\oddsidemargin 5mm
\newcommand{\NP}[1]{Nucl.\ Phys.\ {\bf #1}}
\newcommand{\PL}[1]{Phys.\ Lett.\ {\bf #1}}
\newcommand{\NC}[1]{Nuovo Cimento {\bf #1}}
\newcommand{\CMP}[1]{Comm.\ Math.\ Phys.\ {\bf #1}}
\newcommand{\PR}[1]{Phys.\ Rev.\ {\bf #1}}
\newcommand{\PRL}[1]{Phys.\ Rev.\ Lett.\ {\bf #1}}
\newcommand{\MPL}[1]{Mod.\ Phys.\ Lett.\ {\bf #1}}
\newcommand{\JETP}[1]{Sov.\ Phys.\ JETP {\bf #1}}
\newcommand{\TMP}[1]{Teor.\ Mat.\ Fiz.\ {\bf #1}}

\renewcommand{\thefootnote}{\fnsymbol{footnote}}

\newpage
\setcounter{page}{0}
\begin{titlepage}
\begin{flushright}
\end{flushright}
\begin{center}
{\large  Embedding integrable spin models in solvable vertex models on the square lattice.} \\
\vspace{0.5cm}
{\large M.J. Martins } \\
\vspace{0.15cm}
{\em Universidade Federal de S\~ao Carlos\\
Departamento de F\'{\i}sica \\
C.P. 676, 13565-905, S\~ao Carlos (SP), Brazil\\}
\vspace{0.35cm}
\end{center}
\begin{abstract}
Exploring a mapping among $n$-state spin and vertex models 
on the square lattice, we argue that 
a given integrable spin model with edge weights satisfying the rapidity
difference property can be formulated in the framework 
of an equivalent solvable vertex model.
The Lax operator and the $\mathrm{R}$-matrix associated to the vertex model
are built in terms of the edge weights 
of the spin model
and these operators are shown to satisfy the Yang-Baxter algebra. 
The unitarity of
the $\mathrm{R}$-matrix follows from an assumption that the vertical edge weights
of the spin model satisfies certain local identities known as inversion relation. 
We apply this  embedding to the scalar 
$n$-state Potts model and
we argue that the corresponding $\mathrm{R}$-matrix can be written 
in terms of the underlying
Temperley-Lieb operators. We also consider our construction for the integrable Ashkin-Teller model and the respective $\mathrm{R}$-matrix 
is expressed
in terms of sixteen distinct weights parametrized by theta functions. We 
comment on the possible extension of our results to spin models
whose edge weights are not expressible in terms of the difference of
spectral parameters.
\end{abstract}
\centerline{Keywords: Spin and vertex models, Integrability, Yang-Baxter equation }
\centerline{January~~2025}
\end{titlepage}


\pagestyle{empty}

\newpage

\pagestyle{plain}
\pagenumbering{arabic}

\renewcommand{\thefootnote}{\arabic{footnote}}
\newtheorem{proposition}{Proposition}
\newtheorem{pr}{Proposition}
\newtheorem{remark}{Remark}
\newtheorem{re}{Remark}
\newtheorem{theorem}{Theorem}
\newtheorem{theo}{Theorem}

\def\ll{\left\lgroup}
\def\rr{\right\rgroup}

\newtheorem{Theorem}{Theorem}[section]
\newtheorem{Corollary}[Theorem]{Corollary}
\newtheorem{Proposition}[Theorem]{Proposition}
\newtheorem{Conjecture}[Theorem]{Conjecture}
\newtheorem{Lemma}[Theorem]{Lemma}
\newtheorem{Example}[Theorem]{Example}
\newtheorem{Note}[Theorem]{Note}
\newtheorem{Definition}[Theorem]{Definition}

\section{Introduction}

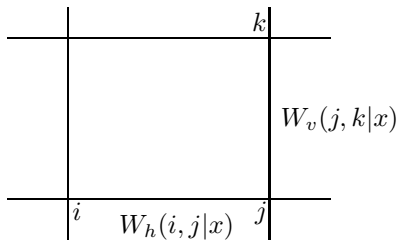
\begin{figure}[ht]
\setlength{\unitlength}{2500sp}
\begin{center}
\begin{picture}(8000,2000)
{\put(1800,1500){\line(1,0){3200}}}
{\put(1800,-100){\line(1,0){3200}}}
{\put(2400,1800){\line(0,-1){2300}}}
{\put(4400,1800){\line(0,-1){2300}}}
{\put(2490,-220){\makebox(0,0){\fontsize{10}{10}\selectfont $i$}}}
{\put(4320,-230){\makebox(0,0){\fontsize{10}{10}\selectfont $j$}}}
{\put(4300,1660){\makebox(0,0){\fontsize{10}{10}\selectfont $k$}}}
{\put(3480,-370){\makebox(0,0){\fontsize{10}{10}\selectfont $W_{h}(i,j|x)$}}}
{\put(5100,690){\makebox(0,0){\fontsize{10}{10}\selectfont $W_{v}(j,k|x)$}}}
\end{picture}
\end{center}
\caption{Schematic representation of the horizontal 
$W_{h}(i,j|x)$
and the vertical 
$W_{v}(j,k|x)$
local Boltzmann weights of the spin models where $(i,j,k)$ indicate spin states.}
\label{FigSpin}
\end{figure}

An important class of integrable two-dimensional lattice systems of 
statistical 
mechanics are models in which the microstate variables or spins are located on the
vertices of the lattice and the interactions occur along the lattice edges \cite{BAX}.
These systems are called spin or edge models and 
the simplest example is a two-state model known as the Ising ferromagnet which was originally 
solved by Onsager 
in 1944 \cite{ONSA1,MCT}. Generalizations of Ising model are achieved considering
that the spins take 
values on a set of integers and 
the energy 
interactions are strictly short-ranged 
depending only on the nearest neighbor's spin states. 
On the square lattice 
the energy interactions among two neighboring spins $(i,j)$ 
can be encoded in terms 
of local horizontal $W_h(i,j|x)$ and vertical 
$W_v(i,j|x)$ edge weights which here we assumed to be 
parametrized by the spectral variable $x$.
The edge weights are schematically shown in Fig.(\ref{FigSpin}) 
where the spin states are considered to take values on a
discrete set $\{1,\dots,n\}$. 

\begin{figure}[ht]
\setlength{\unitlength}{1mm}
\begin{center}
\begin{picture}(55,25)
\put(-36,18){\line(3,-2){16}}
\put(-36,18){\line(-3,-2){16}}
\put(-4,18){\line(3,-2){16}}
\put(-4,18){\line(-3,-2){16}}
\put(15,6){$\cdots$}
\put(15,18){$\cdots$}
\put(25,18){\line(3,-2){16}}
\put(57,18){\line(-3,-2){16}}
\put(57,18){\line(3,-2){16}}
\put(89,18){\line(-3,-2){16}}
\put(89,18){\line(3,-2){16}}
{\put(-51,5){\makebox(0,0){\fontsize{12}{14}\selectfont $a_{1}$}}}
{\put(-19,5){\makebox(0,0){\fontsize{12}{14}\selectfont $a_{2}$}}}
{\put(13,5){\makebox(0,0){\fontsize{12}{14}\selectfont $a_{3}$}}}
{\put(44,5){\makebox(0,0){\fontsize{12}{14}\selectfont $a_{L-1}$}}}
{\put(74,5){\makebox(0,0){\fontsize{12}{14}\selectfont $a_{L}$}}}
{\put(106,5){\makebox(0,0){\fontsize{12}{14}\selectfont $a_{1}$}}}
{\put(-35,20){\makebox(0,0){\fontsize{12}{14}\selectfont $b_{2}$}}}
{\put(-3,20){\makebox(0,0){\fontsize{12}{14}\selectfont $b_{3}$}}}
{\put(28,20){\makebox(0,0){\fontsize{12}{14}\selectfont $b_{L-1}$}}}
{\put(58,20){\makebox(0,0){\fontsize{12}{14}\selectfont $b_{L}$}}}
{\put(90,20){\makebox(0,0){\fontsize{12}{14}\selectfont $b_1$}}}
{\put(-45,16){\makebox(0,0){\fontsize{12}{14}\selectfont $W_{h}$}}}
{\put(-13,16){\makebox(0,0){\fontsize{12}{14}\selectfont $W_{h}$}}}
{\put(48,16){\makebox(0,0){\fontsize{12}{14}\selectfont $W_{h}$}}}
{\put(80,16){\makebox(0,0){\fontsize{12}{14}\selectfont $W_{h}$}}}
{\put(-30,10){\makebox(0,0){\fontsize{12}{14}\selectfont $W_{v}$}}}
{\put(2,10){\makebox(0,0){\fontsize{12}{14}\selectfont $W_{v}$}}}
{\put(32,10){\makebox(0,0){\fontsize{12}{14}\selectfont $W_{v}$}}}
{\put(63,10){\makebox(0,0){\fontsize{12}{14}\selectfont $W_{v}$}}}
{\put(96,10){\makebox(0,0){\fontsize{12}{14}\selectfont $W_{v}$}}}
\end{picture}
\end{center}
\caption{Schematic representation of the diagonal-to-diagonal transfer matrix of spin models.}
\label{FigTDIA}
\end{figure}
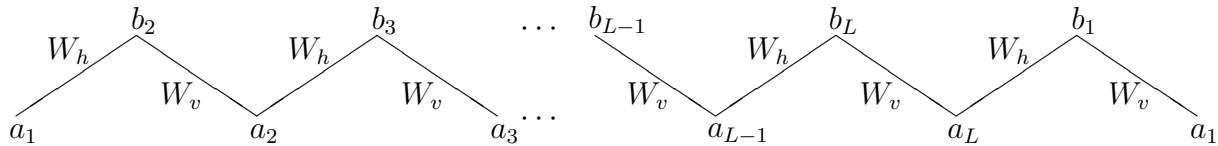

One of the most successful techniques in solving two-dimensional lattice models  
turns out to be the method of commuting transfer matrices \cite{BAX}. The use of transfer matrices 
to solve two-dimensional lattice models originates in the work of Kramers and Wannier \cite{KM} who have shown that 
the respective partition functions can be 
rewritten as the trace of an ordered product of transfer matrices.
For spin models on the square lattice a convenient transfer matrix operator 
is built by considering the product of the weights along a diagonal layer of the square lattice \cite{ONSA2,MIT}. 
The matrix elements
of the diagonal-to-diagonal transfer matrix for a lattice of size $L$ with periodic boundary conditions are,
\begin{equation}
\label{TDIA}
\left[T_{\mathrm{dia}}(x)\right]_{a_1,\dots,a_L}^{b_1,\dots,b_L}= W_v(a_1,b_1 |x) W_h(a_1,b_{2}|x)
W_v(a_2,b_2|x) W_h(a_2,b_{3}|x) \dots 
W_v(a_L,b_L|x) W_h(a_L,b_{1}|x) , 
\end{equation}
which have been schematically illustrated in Fig.(\ref{FigTDIA}). 

It follows that the spin model partition function 
$Z_{\mathrm{spin}}(L)$ on the square lattice of size $L \times L$
can be written in terms of the following trace, 
\begin{equation}
\label{Zspin}
Z_{\mathrm{spin}}(L)= \mathrm{Tr}_{{\cal V}} \left[ \Big(T_{\mathrm{dia}}(x)\Big)^{L} \right] ,
\end{equation}
where ${\cal V}= \displaystyle \prod_{j=1}^{L} \otimes {\mathrm{C}}^{n}$ is usually called
the spin model quantum space.

Given a spin model with weights $W_{h,v}(i,j|x)$  
there exist sufficient conditions for the existence of another spin model with distinct weights 
$W_{h,v}(i,j|y)$  
such that their diagonal-to-diagonal transfer matrices commutes, $[T_{\mathrm{dia}}(x),T_{\mathrm{dia}}(y)]=0$,
for arbitrary lattice sizes.
These are local relations among the horizontal and vertical weights 
originally discovered
in the context of the Ising model \cite{ONSA1,KRA} which are often called
star-triangle
equations, see for instance refs.\cite{PERK,BAX1,PERK1}.  For spin models with weights depending on the
difference of spectral parameters these sets of relations are given by,
\begin{eqnarray}
\label{STAR1}
&& \sum_{d=1}^{n} W_v(d,c|y) W_v(b,d|x-y) W_h(a,d|x)= \mathcal{R}(x,y) W_h(a,b|y) W_h(a,c|x-y) W_v(b,c|x) , \\
\label{STAR2}
&& \sum_{d=1}^{n} W_v(c,d|y) W_v(d,b|x-y) W_h(d,a|x)= \mathcal{R}(x,y) W_h(b,a|y) W_h(c,a|x-y) W_v(c,b|x) ,
\end{eqnarray}
$\mathcal{R}(x,y)$ denotes some factor independent of the spin states 
$a,b,c=1,\dots,n$. 
We remark that many solvable
spin models indeed satisfy Eqs.(\ref{STAR1},\ref{STAR2}) being notable examples  the scalar Potts \cite{POT,BAX3,BASH}, 
the self-dual Ashkin-Teller \cite{ASH,WEG,FEN,SEA},
the Fateev-Zamolodchikov \cite{FAZA}, 
and the Kashiwara-Miwa \cite{KAS,HAS,GAU} models. 

It turns out that recently, we have argued that a $n$-state spin model can be viewed 
as a $n$-state vertex model
in the sense that their partition functions coincide on the finite square lattice 
with periodic boundary conditions \cite{MAR}. In particular, we have exhibited the expression
of the Lax operator encoding the weights of the equivalent vertex
model in terms of a special combination of the horizontal and vertical edge weights
of the spin model. Therefore, assuming that the spin model is also integrable one expects that
the same property should be extended to the equivalent vertex model. In the case of a $n$-state 
vertex model,
integrability is assured by exhibiting an invertible $n^2 \times n^2$ $\mathrm{R}$-matrix 
which together with
the Lax operator has to satisfy a quadratic algebra denominated  Yang-Baxter algebra \cite{BAX2,FAD}. We also remark
that the determination of the $\mathrm{R}$-matrix is an essential ingredient to formulate the exact solution
of the vertex model in terms of the quantum inverse scattering method \cite{FAD}.
In this work, we 
argue that the matrix elements of the $\mathrm{R}$-matrix can be constructed from the spin model edge weights and the respective Yang-Baxter algebra follows from the 
star-triangle equations (\ref{STAR1},\ref{STAR2}). More precisely, we shall show that 
the expression of the underlying 
the $\mathrm{R}$-matrix is,
\begin{equation}
\label{RMA}
\mathrm{R}_{12}(x,y) =\sum_{i,j,k=1}^{n} \frac{W_h(j,i|x) W_v(j,k|x-y)}{W_h(k,i|y)} e_{ik} \otimes e_{ji} ,
\end{equation}
where $e_{i,j}$ denotes the $n \times n$ matrix with only one
non-vanishing entry with value 1 at row
$i$ and column $j$. We note that the $\mathrm{R}$-matrix is not given solely in terms of the difference
of the spectral parameters.
This fact can be explored to generate generalizations of the one-dimensional quantum spin chains underlying the classical lattice spin models.

We have organized this paper as follows. In the next section, we elaborate on our previous results 
concerning the mentioned mapping among $n$-state spin and vertex models 
on the square lattice \cite{MAR}. Here we note that Lax operator of
the equivalent vertex model can be decomposed in terms of the product of two operators depending
either on the horizontal or vertical edge weights. In section 3
we formulate the
Yang-Baxter algebra and show that this algebra is satisfied 
as a consequence of the star-triangle relations for the spin model edge weights. We also discuss certain
properties of the $\mathrm{R}$-matrix such as the unitarity relation and the Yang-Baxter equation.
In section 4 we apply our results for one of the simplest integrable $n$-state spin models which is the scalar Potts model. We argue that the respective $\mathrm{R}$-matrix of the equivalent vertex model
can be rewritten in terms of Temperley-Lieb operators \cite{TEMP}. In section 5 we consider our construction for
the integrable Ashkin-Teller model and show that the corresponding $\mathrm{R}$-matrix 
elements can be expressed in terms of sixteen distinct weights. In section 6 we present our conclusion and comment on the
possibility of extending our embedding to include integrable spin models whose edge weights 
are not parametrized in terms of the difference of spectral parameters. In Appendix A
we present some technical details about the Yang-Baxter equation omitted in the main text.

\section{The spin-vertex equivalence}

In vertex models, the state variables are assigned to 
the links of the square lattice and the Boltzmann weights 
depend on four spin variables $i,j,k,l$ meeting together at the vertex. We express these weights by $w(i,k;j,l|x)$ where the variable $x$ represent the spectral parameter 
as illustrated in Fig.(\ref{FigVer}). 
\setlength{\unitlength}{2500sp}
\begin{figure}[ht]
\begin{center}
\begin{picture}(8000,2000)
{\put(3560,900){\line(1,0){1900}}}
{\put(4500,1800){\line(0,-1){1900}}}
{\put(3430,900){\makebox(0,0){\fontsize{10}{10}\selectfont $i$}}}
{\put(5610,900){\makebox(0,0){\fontsize{10}{10}\selectfont $k$}}}
{\put(4480,-220){\makebox(0,0){\fontsize{10}{10}\selectfont $j$}}}
{\put(4500,1960){\makebox(0,0){\fontsize{10}{10}\selectfont $l$}}}
{\put(2200,900){\makebox(0,0){\fontsize{10}{10}\selectfont $w(i,k;j,l|x)$}}}
\end{picture}
\end{center}
\caption{The 
local Boltzmann weights of vertex models with spin variables $i,j,k,l$ taken values
on the set $\{1,\dots,n\}$.}
\label{FigVer}
\end{figure}
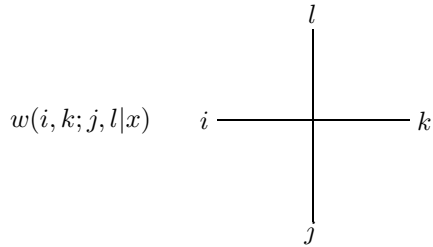

\begin{figure}[ht]
\setlength{\unitlength}{1mm}
\begin{center}
\begin{picture}(55,25)
\put(-38,18){\line(1,0){66}}
\put(-23,10){\line(0,1){16}}
\put(-5,10){\line(0,1){16}}
\put(13,10){\line(0,1){16}}
\put(31,17){$\cdots$}
\put(37,18){\line(1,0){48}}
\put(51,10){\line(0,1){16}}
\put(69,10){\line(0,1){16}}
{\put(-42,18){\makebox(0,0){\fontsize{12}{14}\selectfont $\displaystyle \sum_{c_1,\dots,c_L=1}^{n}$}}}
{\put(-22,8){\makebox(0,0){\fontsize{12}{14}\selectfont $a_1$}}}
{\put(-4,8){\makebox(0,0){\fontsize{12}{14}\selectfont $a_2$}}}
{\put(14,8){\makebox(0,0){\fontsize{12}{14}\selectfont $a_3$}}}
{\put(54,8){\makebox(0,0){\fontsize{12}{14}\selectfont $a_{L-1}$}}}
{\put(70,8){\makebox(0,0){\fontsize{12}{14}\selectfont $a_{L}$}}}
{\put(-22,28){\makebox(0,0){\fontsize{12}{14}\selectfont $b_1$}}}
{\put(-4,28){\makebox(0,0){\fontsize{12}{14}\selectfont $b_2$}}}
{\put(14,28){\makebox(0,0){\fontsize{12}{14}\selectfont $b_3$}}}
{\put(54,28){\makebox(0,0){\fontsize{12}{14}\selectfont $b_{L-1}$}}}
{\put(70,28){\makebox(0,0){\fontsize{12}{14}\selectfont $b_{L}$}}}
{\put(-32,20){\makebox(0,0){\fontsize{12}{14}\selectfont $c_1$}}}
{\put(-14,20){\makebox(0,0){\fontsize{12}{14}\selectfont $c_2$}}}
{\put(4,20){\makebox(0,0){\fontsize{12}{14}\selectfont $c_3$}}}
{\put(22,20){\makebox(0,0){\fontsize{12}{14}\selectfont $c_4$}}}
{\put(42,20){\makebox(0,0){\fontsize{12}{14}\selectfont $c_{L-1}$}}}
{\put(60,20){\makebox(0,0){\fontsize{12}{14}\selectfont $c_{L}$}}}
{\put(78,20){\makebox(0,0){\fontsize{12}{14}\selectfont $c_{1}$}}}
\end{picture}
\end{center}
\caption{Schematic representation of the row-to-row transfer matrix of vertex models.}
\label{FigTVER}
\end{figure}
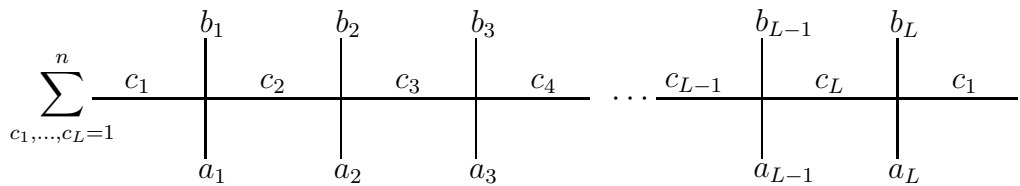

For the vertex models with periodic boundary conditions, it is convenient to build the transfer
matrix considering the product of weights defined 
by two successive rows of spin states of the lattice.
The matrix elements of the row-to-row transfer
matrix is given by,
\begin{equation}
\label{TVER}
\left[T_{\mathrm{ver}}(x)\right]_{a_1,\dots,a_L}^{b_1,\dots,b_L}= \sum_{c_1,\dots,c_L}^{n} w(c_1,c_{2};a_1,b_1|x)
w(c_2,c_{3};a_2,b_2|x) \dots
w(c_L,c_{1};a_L,b_L|x),
\end{equation}
which is schematically illustrated in Fig.(\ref{FigTVER}). 
\vspace{0.5cm}

Once again the vertex model partition function $Z_{\mathrm{ver}}(L)$ on the
$L \times L$ square lattice is given by the 
trace of the  $L$-th power of
the row-to-row transfer matrix,
\begin{equation}
\label{Zver1}
Z_{\mathrm{ver}}(L)= \mathrm{Tr}_{{\cal V}} \left[ \Big(T_{\mathrm{ver}}(x)\Big)^{L} \right] .
\end{equation}

As argued in ref.\cite{MAR} the partition functions 
of the $n$-state spin (\ref{Zspin}) and vertex (\ref{Zver1}) 
models coincide for a finite square lattice providing that 
the corresponding weights satisfy the following relation, 
\begin{equation}
\label{REL}
w(i,k;j,l|x)= W_v(j,k|x) W_h(j,i|x) \delta_{i,l} .
\end{equation}

In fact, by substituting the vertex weights (\ref{REL})  in
the row-to-row transfer matrix (\ref{TVER}) and
after summing over the Kronecker's delta symbols we obtain,
\begin{equation}
\label{TDIAR}
\left[T_{\mathrm{ver}}(x)\right]_{a_1,\dots,a_L}^{b_1,\dots,b_L}= W_h(a_1,b_1|x) W_v(a_1,b_{2}|x)
W_h(a_2,b_2|x) W_v(a_2,b_{3}|x) \dots 
W_h(a_L,b_L|x) W_v(a_L,b_{1}|x) , 
\end{equation}
and we note that these matrix elements have the basic form of those 
of the diagonal-to-diagonal 
transfer matrix (\ref{TDIA}) 
except by the fact that the horizontal and vertical edge weights are interchanged.
However, this exchange among the weights corresponds to the rotation of the lattice by $90^{0}$ degrees, and such operation does not affect the spin model partition function on the square lattice with periodic boundary conditions.

An interesting feature of the vertex models is that they have an inherent
tensor structure. The Boltzmann weights can be encoded in terms of a local vertex operator usually called the Lax operator 
which acts on the
tensor product of two spaces associated to the horizontal and 
vertical spin variables. It turns out that the Lax operator can be written as follows,
\begin{equation}
\label{LAX}
\mathbb{L}_{12}(x)= \sum_{i,j,k,l=1}^{n} w(i,k;j,l|x) e_{ik} \otimes e_{jl}
=\sum_{i,j,k=1}^{n} W_h(j,i|x) W_v(j,k|x) e_{ik} \otimes e_{ji} ,
\end{equation}
and as a result, the respective row-to-row transfer matrix can be compactly expressed in terms of
trace over an auxiliary space $\mathcal{A} \equiv C^{n}$, namely
\begin{equation}
\label{TRAN}
T_{\mathrm{ver}}(x)= \mathrm{Tr}_{\cal A} \left[ \mathbb{L}_{{\cal A}1}(x) \mathbb{L}_{{\cal A}2}(x) \dots \mathbb{L}_{{\cal A}L}(x) \right] .
\end{equation}

In our case, we observe that the Lax operator can be decomposed 
in terms of the product of two operators whose entries depend either on the 
horizontal or on the vertical spin edge weights. In fact, it is possible 
to the rewrite 
the Lax operator as 
$\mathbb{L}_{12}(x)=\mathbb{L}^{(h)}_{12}(x) \mathbb{L}^{(v)}_{12}(x)$ such that 
the components are given by,
\begin{equation}
\mathbb{L}^{(h)}_{12}(x)=\sum_{i,j=1}^{n} W_h(j,i|x) e_{ii} \otimes e_{jj},~~~ 
\mathbb{L}^{(v)}_{12}(x)=\sum_{i,j,k=1}^{n} W_v(j,k|x) e_{ik} \otimes e_{ji} , 
\end{equation}
where we note that the operator associated to the horizontal weights is a diagonal matrix.

We next would like to discuss the Hamiltonian limit associated with the $n$-state spin or vertex models. 
To this end, we assume the existence of a particular spectral parameter point, say $x=0$,
in which the edge weights satisfy the
following initial conditions,
\begin{equation}
\label{BOUND}
W_h(i,j|0)=1,~~~W_v(i,j|0)= \delta_{i,j} ,
\end{equation}
implying that at $x=0$ the Lax operator becomes the permutator on the 
tensor product $C^{n} \otimes C^{n}$,
\begin{equation}
\label{INI}
\mathbb{L}_{12}(0)= P_{12}=\displaystyle \sum_{i,j=1}^{n} e_{ij} \otimes e_{ji} .
\end{equation}

Taking into account the condition (\ref{INI}) the corresponding quantum spin chain
is obtained by taking the logarithmic derivative of the row-to-row transfer matrix (\ref{TRAN})
at the special point $x=0$ \cite{BAX2}. This leads us to the following Hamiltonian,
\begin{equation}
\label{HAM}
H= \sum_{j=1}^{L} \left[
\sum_{i,k=1}^{n} \frac{d}{dx} [W_h(i,k|x)]\Big{|}_{x=0} e_{ii}^{(j)} \otimes e_{kk}^{(j+1)} 
+\sum_{i,k,l=1}^{n}\frac{d}{dx} [W_v(i,k|x)]\Big{|}_{x=0} e_{ik}^{(j)} \otimes e_{ll}^{(j+1)} \right] ,
\end{equation}
where $e_{ik}^{(j)}$ denotes the action of the matrix $e_{ik}$ on the $j$-th site of a chain 
of size $L$ and
periodic boundary condition is imposed by defining
$e_{ik}^{(L+1)}=e_{ik}^{(1)}$. 

We would like to conclude by mentioning certain local identities satisfied by edge weights of solvable spin models such as 
the scalar Potts \cite{POT,BAX3,BASH}, 
the self-dual Ashkin-Teller \cite{ASH,WEG,FEN,SEA},
the Fateev-Zamolodchikov \cite{FAZA}, 
and the Kashiwara-Miwa \cite{KAS,HAS,GAU} models. 
These identities are usually called 
inversion relations \cite{BAX}
and their expressions are given by,
\begin{equation}
\label{INV}
W_h(i,j|x)W_h(i,j|-x)=\rho_1(x),~~~\sum_{k=1}^{n} W_v(i,k|x) W_v(k,j|-x)= \rho_2(x) \delta_{i,j} ,
\end{equation}
where $\rho_{1,2}(x)$ are arbitrary auxiliary normalization factors.

In the next section, we argue that the inversion
relation associated with the vertical edge weights is relevant to setting up the unitarity property 
of the proposed $\mathrm{R}$-matrix (\ref{RMA}).
This feature ensures that the underlying $\mathrm{R}$-matrix 
is invertible for almost all spectral parameters $x$ and $y$.

\section{The Yang-Baxter algebra}

If the spin model is integrable, it is natural to expect that this property will be
also satisfied by the equivalent vertex model. A sufficient condition for commuting transfer matrices of vertex models was first introduced by Baxter in his analysis of the eight-vertex
model \cite{BAX2} and afterward elaborated in the context of the quantum
inverse scattering approach \cite{FAD}. This condition requires the existence of an invertible $\mathrm{R}$-matrix
which together with the Lax operators satisfies the Yang-Baxter algebra. This algebra 
involves
the action of these operators on the tensor product of three $n$-dimensional spaces, namely
\begin{equation}
\label{RLLpure}
\mathrm{R}_{12}(x,y) \mathbb{L}_{13}(x) \mathbb{L}_{23}(y)= 
\mathbb{L}_{23}(y) \mathbb{L}_{13}(x) \mathrm{R}_{12}(x,y) ,
\end{equation}
where the lower indices indicate the spaces in which the Lax and the $\mathrm{R}$-matrix operators 
acts non-trivially.

We observe that in the above relation, the Lax operators are fixed 
by the edge weights of the
spin model, see Eq.(\ref{LAX}). In order to show that this algebra is satisfied by 
the $\mathrm{R}$-matrix (\ref{RMA}) we find  convenient to rewrite the
Yang-Baxter algebra (\ref{RLLpure}) in terms of its components. To this end we express
the $\mathrm{R}$-matrix (\ref{RMA}) as,
\begin{equation}
\label{RMA1}
\mathrm{R}_{12}(x,y)= \sum_{i,j,k,l} \mathrm{R}_{i,j}^{k,l}(x,y) e_{ik} \otimes e_{jl},~~~
\mathrm{R}_{i,j}^{k,l}(x,y)=
\frac{W_h(j,i|x) W_v(j,k|x-y)}{W_h(k,i|y)} \delta_{i,l} ,
\end{equation}
and by using the expression of the Lax operator (\ref{LAX}) we find that the 
matrix elements of the Yang-Baxter algebra are,
\begin{eqnarray}
\label{YBAR}
&& \sum_{\gamma=1}^{n} \mathrm{R}_{a_1,a_2}^{\gamma,b_3}(x,y) W_h(a_3,\gamma|x) W_v(a_3,b_1|x) 
W_h(\gamma,b_3|y) W_v(\gamma,b_2|y)= \nonumber \\ 
&& \delta_{a_1,b_3} \sum_{\gamma,\gamma^{'}=1}^{n} 
\mathrm{R}_{\gamma,\gamma^{'}}^{b_1,b_2}(x,y)  
W_h(a_3,a_2|y) W_v(a_3,\gamma^{'}|y) 
W_h(a_2,a_1|x) W_v(a_2,\gamma|x) .
\end{eqnarray}

In what follows, we shall show that the Yang-Baxter algebra (\ref{YBAR}) is indeed fulfilled. We first substitute on the 
left hand side of Eq.(\ref{YBAR}) the structure constants 
$\mathrm{R}_{a_1,a_2}^{\gamma,b_3}(x,y)$ 
and sum over the Kronecker's delta symbols. Further simplification can be carried out by using the first set of the star-triangle
relations (\ref{STAR1}). As a result of these steps we obtain,
\begin{eqnarray}
\label{LEFT}
&& \sum_{\gamma=1}^{n} \mathrm{R}_{a_1,a_2}^{\gamma,b_3}(x,y) W_h(a_3,\gamma|x) W_v(a_3,b_1|y) 
W_h(\gamma,b_3|y) W_v(\gamma,b_2|y)= \nonumber \\ 
&& \delta_{a_1,b_3} W_v(a_3,b_1|x) W_h(a_2,a_1|x) 
\sum_{\gamma=1}^{n}  W_v(\gamma,b_2|y) W_v(a_2,\gamma|x-y) W_h(a_3,\gamma|x) = \nonumber \\
&& \delta_{a_1,b_3} W_v(a_3,b_1|x) W_h(a_2,a_1|x) \mathcal{R}(x,y)
W_h(a_3,a_2|y) W_h(a_3,b_2|x-y)  
W_v(a_2,b_2|x) .
\end{eqnarray}

We now repeat the same procedure explained above for the right 
hand side of Eq.(\ref{YBAR}). Here
we use the second set of the star-triangle relations (\ref{STAR2}) to simplify the sum of the product of edge weights. The steps are summarized below,
\begin{eqnarray}
\label{RIGHT}
&& \delta_{a_1,b_3} \sum_{\gamma,\gamma^{'}=1}^{n} 
\mathrm{R}_{\gamma,\gamma^{'}}^{b_1,b_2}(x,y)  
W_h(a_3,a_2|y) W_v(a_3,\gamma^{'}|y) 
W_h(a_2,a_1|x) W_v(a_2,\gamma|x) = \nonumber \\
&& \delta_{a_1,b_3} \frac{W_h(a_2,a_1|x) W_h(a_3,a_2|y) W_v(a_2,b_2|x)}{W_h(b_1,b_2|y)}
\sum_{\gamma^{'}=1}^{n}  W_v(a_3,\gamma^{'}|y) W_v(\gamma^{'},b_1|x-y) W_h(\gamma^{'},b_2|x) = \nonumber \\
&& \delta_{a_1,b_3} W_h(a_2,a_1|x) W_h(a_3,a_2|y) W_v(a_2,b_2|x) \mathcal{R}(x,y)  W_h(a_3,b_2|x-y) W_v(a_3,b_1|x) ,
\end{eqnarray}
and by comparing Eqs.(\ref{LEFT},\ref{RIGHT}) we see that the left and the right hand 
sides of the Yang-Baxter algebra 
are the same.

The other important step for commuting transfer matrices is the assumption that the
underlying $\mathrm{R}$-matrix is invertible
for almost all spectral parameters $x$ and $y$. In our case, this feature follows from the fact that the 
$\mathrm{R}$-matrix (\ref{RMA1}) satisfy the unitarity property, 
\begin{equation}
\label{UNI}
\mathrm{R}_{12}(x,y)
\mathrm{R}_{21}(y,x)=\rho_2(x-y) \mathrm{I_n} \otimes \mathrm{I_n},
\end{equation}
where $\mathrm{I}_n$ denotes 
the $n \times n$ identity matrix. It turns out that this property can be derived from direct computation
and with the help of the inversion relation for the vertical edge weights (\ref{INV}), namely   
\begin{eqnarray}
\mathrm{R}_{12}(x,y)
\mathrm{R}_{21}(y,x)&=& \sum_{i,j,l=1}^{n} \frac{W_h(j,i|x)}{W_h(l,i|x)} \sum_{k=1}^{n} \left[W_v(j,k|x-y) W_v(k,l|y-x)\right] e_{ii} \otimes e_{j,l} \nonumber \\
&=&\sum_{i,j,l=1}^{n} \frac{W_h(j,i|x)}{W_h(l,i|x)} \rho_2(x-y) \delta_{j,l}  e_{ii} \otimes e_{j,l} \nonumber \\
&=&\rho_2(x-y) \sum_{i,j=1}^{n} e_{ii} \otimes e_{j,j} =\rho_2(x-y) \mathrm{I}_n \otimes \mathrm{I}_n .
\end{eqnarray}

At this point, we have discussed the basic ingredients showing that the equivalent $n$-state
vertex model associated with an integrable $n$-state spin model indeed gives rise to a family of 
commuting row-to-row transfer matrices. In the context of vertex models is assumed that
the Yang-Baxter algebra is an associative algebra when you reorder 
the product of three Lax operators with distinct rapidities $x$, $y$ and $z$. 
It is well known that 
a sufficient condition for the associativity property
is the celebrated Yang-Baxter equation,
\begin{equation}
\label{YBE}
\mathrm{R}_{12}(x,y)
\mathrm{R}_{13}(x,z)
\mathrm{R}_{23}(y,z)=
\mathrm{R}_{23}(y,z)
\mathrm{R}_{13}(x,z)
\mathrm{R}_{12}(x,y) ,
\end{equation}
where at $z=0$ reduces to the Yang-Baxter algebra (\ref{RLLpure}) since we have 
the identity $\mathrm{R}_{ab}(x,0)= \mathbb{L}_{ab}(x)$ as a consequence
of the initial condition (\ref{BOUND}). 

It is plausible to think that the Yang-Baxter equation (\ref{YBE}) 
should follow from 
systematic applications of the star-triangle equations (\ref{STAR1},\ref{STAR2}) 
combined with the help of the inversion relations (\ref{INV}). In this sense,
we remark that it has been argued that a particular combination of four edge weights of
a given solvable spin model one can obtain the Boltzmann weights of 
a vertex model with commuting 
transfer matrices \cite{PERK1,MCPERK2,BASTR}. In this construction,
the number of spectral variables of the vertex model weights
is duplicated due to the fact that it is used in two distinct sets
of rapidities to parametrize horizontal and vertical spin edge weights. We now follow the construction of refs.\cite{PERK1,MCPERK2,BASTR} in the situation
where the spin edge weights depend only on the difference 
of the spectral parameters. In our notation, the 
matrix elements of the Lax operator associated to such vertex model are,
\begin{equation}
\label{RMAGE}
\mathbb{\widetilde{L}}_{i,j}^{k,l}(x_1;x_2,y_1;y_2)= W_h(i,j|x_1-y_1) W_v(j,k|x_1-y_2) W_v(i,l|x_2-y_1) W_h(l,k|x_2-y_2)
\end{equation}
where $x_1,y_1$ and $x_2,y_2$ denote two pairs of rapidities. 

We next apply the expression (\ref{RMAGE}) in a particular case  of rapidities arrangements by choosing $x_2=y_1$, $x_1 =y_1+x$ and $y_2=y_1+y$ where $x$ and $y$
are arbitrary variables. By using
the initial condition (\ref{BOUND}) for the vertical edge weights 
as well as the inversion relation (\ref{INV}) for the horizontal 
edge weights we obtain,
\begin{eqnarray}
\mathbb{\widetilde{L}}_{i,j}^{k,l}(y_1+x;y_1,y_1;y_1+y) &=& W_h(i,j|x) W_v(j,k|x-y) W_h(l,k|-y) \delta_{i,l} \nonumber \\
&=& \rho_1(y) \frac{W_h(i,j|x) W_v(j,k|x-y)} {W_h(l,k|y)} \delta_{i,l} 
\end{eqnarray}

By comparing the above elements with the entries of the 
proposed $\mathrm{R}$-matrix (\ref{RMA1}) we note that they are similar
except by the fact that the spin states
of the horizontal edge weights are exchanged. At this point, we recall
that the most known solvable
spin models with weights parametrized by the difference in the spectral
parameters have
the reflection symmetry $W_{h,v}(i,j|x)=W_{h,v}(j,i|x)$. Therefore, we expect
that the $\mathrm{R}$-matrices (\ref{RMA1}) associated to the 
scalar Potts,  
the self-dual Ashkin-Teller 
the Fateev-Zamolodchikov, 
and the Kashiwara-Miwa spin models should indeed satisfy the Yang-Baxter equation (\ref{YBE}). 
In fact, we have confirmed this relation 
for the above mentioned  spin models by using the explicit expressions of the corresponding edge weights. The technical details 
concerning this verification have been summarized in Appendix A.  
We also remark that recently it has been argued that under certain conditions any $\mathrm{R}$-matrix satisfying 
a given Yang-Baxter algebra is a solution of
the Yang-Baxter equation \cite{LEPO}.

The validity of the Yang-Baxter equation for a given $\mathrm{R}$-matrix which is not of
difference form has the following interesting consequence. We can formulate a 
generalized integrable $n$-state vertex
model by replacing the Lax operator by the $\mathrm{R}$-matrix in the row-to-row transfer matrix,
\begin{equation}
\label{TRAG}
T(x,x_0)= \mathrm{Tr}_{\cal A} \left[ \mathrm{R}_{{\cal A}1}(x,x_0) \mathrm{R}_{{\cal A}2}(x,x_0) \dots \mathrm{R}_{{\cal A}L}(x,x_0) \right] ,
\end{equation}
where the second spectral parameter $x_0$ plays the role
of an additional independent coupling of the model. We note that for $x_0=0$ we recover 
the transfer matrix of the $n$-state vertex model
equivalent to a $n$-state spin model, see Eq.(\ref{TRAN})

We next observe that for $x=x_0$ each operator $\mathrm{R}_{{\cal A}j}(x,x_0)$ 
equals the permutator on the corresponding tensor product spaces $C^{n} \otimes C_j^{n}$. We can therefore
construct a generalized quantum spin chain by taking the logarithmic derivative of the transfer matrix (\ref{TRAG}) at the point $x=x_0$. The expression of the 
respective Hamiltonian is,
\begin{eqnarray}
H(x_0) &=& \sum_{j=1}^{L} 
\sum_{i,k=1}^{n} \left[\frac{1}{W_h(i,k|x_0)}\frac{d}{dx} [W_h(i,k|x)]\Big{|}_{x=x_0} e_{ii}^{(j)} \otimes e_{kk}^{(j+1)} \right] 
\nonumber \\
&+& \sum_{j=1}^{L} \left[ \sum_{i,k,l=1}^{n} \frac{W_h(i,l|x_0)}{W_h(k,l|x_0)}\frac{d}{dx} [W_v(i,k|x-x_0)]\Big{|}_{x=x_0} e_{ik}^{(j)} \otimes e_{ll}^{(j+1)} \right] , 
\end{eqnarray}
where periodic boundary condition is assumed. For $x_0 \neq 0$ the above Hamiltonian generalizes 
the quantum spin chain associated to the lattice $n$-state spin or vertex models, see Eq.(\ref{HAM}).

In the next sections, we apply the above results in the cases of the solvable scalar Potts  
and the Ashkin-Teller spin models.

\section{The scalar Potts model}

The $n$-state scalar Potts model is a generalization of the Ising model 
when at each $i$-th site
the spin variables $\sigma_i$ can 
have $n \geq 2$ possible values. It is assumed that 
the interactions among the horizontal (vertical)
adjacent spins 
variables have the same thermal energy $J_h$ ($J_v$) 
when the respective spins variables are alike 
and zero if they
are different \cite{POT}. The total thermal energy 
of this spin model
is given by,
\begin{equation}
\frac{E}{k_{B} T}= -J_v \sum_{<i,j>)} \delta(\sigma_i,\sigma_j) 
-J_v \sum_{<k,l>} \delta(\sigma_k,\sigma_l) ,
\end{equation}
where $k_{B}$ is Boltzmann's constant and $T$ is the temperature. 
The symbols $<i,j>$ and $<k,l>$ indicate the summations over all the 
horizontal and vertical
edges of the square lattice, respectively.  

This means that  the edge weights have only two basic elements
and they can be written as follows,
\begin{equation}
W_h(i,j)= \kappa_v \left[ (\exp(J_h)-1)\delta_{i,j}+1 \right],~~
W_v(i,j)= \kappa_h \left[ (\exp(J_v)-1)\delta_{i,j}+1 \right] ,
\end{equation}
where $\kappa_{h,v}$ are normalization factors at our disposal. This model cannot be solved in general,
but it is integrable when the couplings sit on the self-dual manifold \cite{BAX3},
\begin{equation}
\left(\exp(J_h)-1\right)
\left(\exp(J_v)-1\right)=n .
\end{equation}

It turns out that 
one possible parametrization of such a solvable manifold is
given as follows \cite{BAX3,BASH},
\begin{equation}
W_h(i,j|x)= 1 +\sqrt{n} f_n(x) \delta_{i,j},~~
W_v(i,j|x)= \frac{f_n(x)}{\sqrt{n}} +\delta_{i,j} ,
\end{equation}
where the function $f_n(x)$ is defined by,
\begin{equation}
f_n(x)=\begin{cases}
\frac{\sin(x)}{\sin(\gamma_n-x)} & \mathrm{for}~~ n=2,3 \\
\frac{x}{\gamma_n-x} & \mathrm{for}~~ n=4 \\
\frac{\sinh(x)}{\sinh(\gamma_n-x)} & \mathrm{for}~~ n \geq 5 \\
\end{cases},~~~
\gamma_n=\begin{cases}
\arccos\left(\frac{\sqrt{n}}{2} \right)& \mathrm{for}~~ n=2,3 \\
1 & \mathrm{for}~~ n=4 \\
\arccosh\left(\frac{\sqrt{n}}{2} \right)& \mathrm{for}~~ n \geq 5 \\
\end{cases}
\end{equation}
and the normalization functions of the inversion
relations are trivial,
i.e $\rho_1(x)=\rho_2(x)=1$.

The above parametrization can be traced back to the fact that the $n$-state
Potts model can be seen as one of the possible representations of the 
Temperley-Lieb algebra \cite{TEMP}. In this context,
the function $f_n(x)$ arises as a solution of certain functional 
equation associated to the Baxterization of a braid originated from the 
Temperley-Lieb monoid \cite{JON}. For the scalar 
Potts model
the corresponding Temperley-Lieb  generators
can be expressed in terms of the basic elements
of the $Z(n)$ algebra,
\begin{equation}
Z^{n}=X^{n}=1,~~ZX=\omega XZ ,
\end{equation}
where $\omega=\exp(2\pi i/n)$ and the entries of the $n \times n$ matrices $Z$ and $X$ are,
\begin{equation}
Z_{k,l}= \omega^{k-1} \delta_{k,l},~~~X_{k,l}=\delta_{k,l+1}~(\mathrm{mod}~n) .
\end{equation}

By now the $n$-state scalar Potts representation of the Temperley-Lieb algebra is well known in the literature, see for instance \cite{BERK}. In terms of
the $Z(n)$ operators such representation is given by, 
\begin{equation}
\label{TLGEN}
E_{2j}= \frac{1}{\sqrt{n}} \sum_{k=0}^{n-1} \left( Z_j  Z^{\dagger}_{j+1} \right)^{k},~~
E_{2j-1}= \frac{1}{\sqrt{n}} \sum_{k=0}^{n-1} \left(X_j\right)^{k} ,
\end{equation}
where the operators $Z_j$ and $X_j$ acting at the $j$-th site of the chain obey the $Z(n)$ algebra.
These generators satisfy the following algebraic relations,
\begin{equation}
\left(E_j\right)^2= \sqrt{n} E_j,~~E_j E_{j \pm 1} E_j=E_j,~~[E_j,E_k]=0~\mathrm{for}~|l-k| \geq 2 .
\end{equation}

From the above discussion, we see that for the scalar Potts model the most natural operators to use 
are the Temperley-Lieb generators (\ref{TLGEN}). Therefore, one expects that 
both the Lax operator of the equivalent vertex model 
and the respective $\mathrm{R}$-matrix satisfying the Yang-Baxter algebra should be 
rewritten in terms of the Temperley-Lieb generators. In fact, the expression for the 
Lax operator (\ref{LAX}) is,
\begin{equation}
\mathbb{L}_{12}(x)=P_{12} \Big( \mathrm{I}_n \otimes \mathrm{I_n}+f_n(x)E_2 \Big)
\Big( \mathrm{I}_n \otimes \mathrm{I_n}+f_n(x)E_1 \Big) ,
\end{equation}
while the corresponding $\mathrm{R}$-matrix (\ref{RMA1}) is given by,
\begin{equation}
\label{POTRMA}
\mathrm{R}_{12}(x,y)=P_{12} \Big( \mathrm{I}_n \otimes \mathrm{I_n}+f_n(x-y)\left[E_1+E_2 +f_n(x)E_2E_1+f_n(-y)E_1E_2\right] \Big) .
\end{equation}

By using trigonometric identities and the properties of the 
Temperley-Lieb operators 
one can verify that  the $\mathrm{R}$-matrix (\ref{POTRMA}) 
indeed satisfy the Yang-Baxter equation.  Interestingly enough, this result tells us 
that a particular combination of two different Temperley-Lieb operators appears to be suitable
for Baxterization once we consider that the 
$\mathrm{R}$-matrix can not be parametrized
only in terms of the difference in the rapidities. As explained at the end of the previous  section we can use this $\mathrm{R}$-matrix to generate a generalized $Z(N)$ invariant quantum spin chain and as usual, we write the respective 
Hamiltonian as,
\begin{equation}
\label{HAMG1}
H(x_0)= -J\Big[ \sum_{j=1}^{L-1} H_{j,j+1}(x_0)  +H_{L,1}(x_0) \Big] ,
\end{equation}
where $J$ is an overall normalization. 
The corresponding two-body Hamiltonian $H_{j,j+1}$, apart from an additive term, is given
in terms of the $Z(n)$ operators by the following 
expression\footnote{We observe that for $n=2,3$ we
recover the spin chains discussed in our previous work \cite{MAR}.} ,
\begin{equation}
\label{HAMG2}
H_{j,j+1}(x_0)= \sum_{k=1}^{n-1} (X_j)^k +
\sum_{k=1}^{n-1} (Z_j Z^{\dagger}_{j+1})^k 
+g_n(x_0) \sum_{k,l=1}^{n-1} (Z_j Z^{\dagger}_{j+1})^k (X_j)^l
+g_n(-x_0) \sum_{k,l=1}^{n-1} (X_j)^k (Z_j Z^{\dagger}_{j+1})^l , 
\end{equation}
where the function $g_n(x)$ is given by,
\begin{equation}
\label{HAMG3}
g_n(x)=\begin{cases}
\frac{\sin(x) \sin(\gamma_n+x)}{\sin(2\gamma_n) \sin(\gamma_n)} & \mathrm{for}~~ n=2,3 \\
\frac{x(1+x)}{2} & \mathrm{for}~~ n=4 \\
\frac{\sinh(x) \sinh(\gamma_n+x)}{\sinh(2\gamma_n) \sinh(\gamma_n)} & \mathrm{for}~~ n \geq 5 .\\
\end{cases}
\end{equation}

The Hamiltonian defined by Eqs.(\ref{HAMG1},\ref{HAMG2},\ref{HAMG3}) 
is an integrable $Z(n)$ symmetric deformation of the quantum spin chain associated with the scalar Potts model. We note that this operator is Hermitian 
when $x_0$ 
is an imaginary number. The first two terms of the Hamiltonian correspond to the standard
spin chain obtained within the so-called time-continuum limit 
of a certain transfer matrix formulation  of the 
classical scalar Potts model \cite{MIT1,SOL} while 
the last two terms are
additional interactions among particular combinations of $Z(n)$ generators. 
We recall that these types of extra interactions have appeared before 
in a quantum spin chain 
derived from the integrable higher-spin $XXZ$ Heisenberg model for 
a specific choice of the 
quantum group deformation parameter \cite{VERN}. We remark, however, 
that in the model of ref.\cite{VERN}
the extra interactions are weighted by suitable factors such that the 
underlying Hamiltonian is $U(1)$-invariant.

\section{The Ashkin-Teller model}

The Ashkin-Teller model \cite{ASH} can be formulated in terms of 
two Ising models with thermal energies $J_{v,h}$ and $K_{v,h}$  which are coupled by a four spin interactions $L_{h,v}$ involving the product of the energy 
terms of the Ising models \cite{FAN}.
If we denote the Ising spins at a given $i$-th site by the variables
$\sigma_i$ and $\tau_i$ taking the values $\pm 1$ the respective
thermal energy can be written as follows,
\begin{equation}
\frac{E}{k_B T}= -\sum_{<i,j>} \Big ( J_h \sigma_i \sigma_j+K_h \tau_i \tau_j +L_h \sigma_i \sigma_j \tau_i \tau_j \Big ) 
-\sum_{<k,l>} \Big ( J_v \sigma_k \sigma_l+K_v \tau_k \tau_l +L_v \sigma_k \sigma_l \tau_k \tau_l \Big ) ,
\end{equation}
where the sums $<i,j>$ and $<k,l>$ are over the horizontal and vertical 
edges of the square lattice.

This means that the Ashkin-Teller 
is a four-state spin model 
and the edge weights 
can be represented by the following $4 \times 4$ matrices,
\begin{equation}
W_h=\left(
\begin{array}{cccc}
a_h & b_h & c_h & d_h \\
b_h & a_h & d_h & c_h \\ 
c_h & d_h & a_h & b_h \\ 
d_h & c_h & b_h & a_h \\ 
\end{array}
\right) ,~~~
W_v=\left(
\begin{array}{cccc}
a_v & b_v & c_v & d_v \\
b_v & a_v & d_v & c_v \\ 
c_v & d_v & a_v & b_v \\ 
d_v & c_v & b_v & a_v \\ 
\end{array}
\right) ,
\end{equation}
where the relation among the edge weights with the couplings are,
\begin{eqnarray}
&& a_h= \kappa_h e^{(J_h+K_h+L_h)},~~b_h= \kappa_h e^{(J_h-K_h-L_h)},~~
c_h= \kappa_h e^{(-J_h+K_h-L_h)},~~d_h=\kappa_h e^{(-J_h-K_h+L_h)}, \nonumber \\
&& a_v= \kappa_v e^{(J_v+K_v+L_v)},~~b_v= \kappa_v e^{(J_v-K_h-L_v)},~~
c_v=\kappa_v e^{(-J_v+K_v-L_v)}~~d_v=\kappa_v e^{(-J_v-K_v+L_v)} ,
\end{eqnarray}
such that $\kappa_{h,v}$ are arbitrary normalizations factors.

It has been argued that the Ashkin-Teller spin model can be converted   into a staggered eight-vertex model 
on the square lattice \cite{WEG} which becomes integrable when the
vertex weights on the two sublattices are proportional \cite{BAX}. If we denote 
by $w_a$, $w_b$, $w_c$ and
$w_d$ the weights of the eight-vertex model it turns out that the edge weights of the 
Ashkin-Teller spin model
are given by \cite{FEN},
\begin{eqnarray}
&& a_h=1,~~b_h=\frac{w_a-w_d}{w_b+w_c},~~
c_h=\frac{w_a+w_d}{w_b+w_c},~~
d_h=\frac{w_c-w_b}{w_b+w_c} \nonumber \\
&& a_v=1,~~b_v=\frac{w_b-w_d}{w_a+w_c},~~
c_v=\frac{w_b+w_d}{w_a+w_c},~~
d_v=\frac{w_c-w_a}{w_a+w_c} ,
\end{eqnarray}
where we have normalized the horizontal and vertical 
edge weights by $a_h$ and $a_v$, respectively.

It is well known that weights of the eight-vertex model can be 
uniformized in terms of the theta elliptic functions \cite{BAX}. 
By using this parametrization one finds that
the spin edge weights can be expressed as follows \cite{SEA},
\begin{eqnarray}
\label{TETAW}
&& a_h(x)=1,~~~~~~~~~~~~~~~~~~~~~~~~~~~~ a_v(x)=1, \nonumber \\
&& b_h(x)=\frac{\theta_1\left(\frac{\xi-x}{2},q \right) \theta_3 \left( \frac{\xi+x}{2},q \right )}
{\theta_3\left(\frac{\xi-x}{2},q \right) \theta_1 \left( \frac{\xi+x}{2},q\right )},~~
b_v(x)=\frac{\theta_1\left(\frac{x}{2},q \right) \theta_3 \left( \xi-\frac{x}{2},q \right )}
{\theta_3 \left(\frac{x}{2},q \right) \theta_1 \left( \xi- \frac{x}{2},q \right )}, \nonumber \\
&& c_h(x)=\frac{\theta_1\left(\frac{\xi-x}{2},q \right) \theta_4 \left( \frac{\xi+x}{2},q \right )}
{\theta_4\left(\frac{\xi-x}{2},q \right) \theta_1 \left( \frac{\xi+x}{2}, q\right )},~~
c_v(x)=\frac{\theta_1\left(\frac{x}{2},q \right) \theta_4 \left( \xi-\frac{x}{2}, q \right )}
{\theta_4 \left(\frac{x}{2},q \right) \theta_1 \left( \xi-\frac{x}{2},q \right )},  \nonumber \\
&& d_h(x)=\frac{\theta_1\left(\frac{\xi-x}{2},q \right) \theta_2 \left( \frac{\xi+x}{2}, q \right )}
{\theta_2\left(\frac{\xi-x}{2},q \right) \theta_1 \left( \frac{\xi+x}{2},q \right )},~~
d_v(x)=\frac{\theta_1\left(\frac{x}{2},q \right) \theta_2 \left( \xi-\frac{x}{2}, q \right )}
{\theta_2\left(\frac{x}{2},q \right) \theta_1 \left( \xi-\frac{x}{2},q \right )} ,
\end{eqnarray}
where $\xi$ is an arbitrary parameter and $\theta_i(x,q)$ $i=1,\dots,4$ are
the four standard theta functions   
of nome $q$, with $|q|<1$, defined by \cite{GRA},
\begin{eqnarray}
\theta_1(x,q)&=& 2 q^{1/4} \sin(x) \prod_{k=1}^{\infty} (1-2 q^{2k} \cos(2x)+q^{4k})(1-q^{2k}) \nonumber \\
\theta_2(x,q)&=& 2 q^{1/4} \cos(x) \prod_{k=1}^{\infty} (1+2 q^{2k} \cos(2x)+q^{4k})(1-q^{2k}) \nonumber \\
\theta_3(x,q)&=& \prod_{k=1}^{\infty} (1+2 q^{2k-1} \cos(2x)+q^{4k-2})(1-q^{2k}) \nonumber \\
\theta_4(x,q)&=& \prod_{k=1}^{\infty} (1-2 q^{2k-1} \cos(2x)+q^{4k-2})(1-q^{2k}) ,
\end{eqnarray}
while the normalizations 
entering the inversion relations are,
\begin{equation}
\rho_1(x)=1,~~ \rho_2(x)= 4 \Bigg[ \frac{\theta_1(\frac{x}{2},q)}{\theta_1(x,q)} \Bigg]^2 \frac{\theta_1(\xi-x,q) \theta_1(\xi+x,q)}
{\theta_1(\xi-\frac{x}{2},q) \theta_1(\xi+\frac{x}{2},q)} .
\end{equation}

We now use the result (\ref{RMA1}) to built the $\mathrm{R}$-matrix of the equivalent vertex model associated to the 
Ashkin-Teller model. We find that this operator has sixty-four non-null vertex weights but many of them are the same due to 
the underlying $Z(2) \times Z(2)$ symmetry of the spin model. It turns out that we have only sixteen distinct weights 
and the explicit form of the $16 \times 16$ $\mathrm{R}$-matrix is given by,

{\scriptsize
\begin{equation}
\label{RASH}
\mathrm{R}_{12}(x,y)=\left(
\begin{array}{cccc|cccc|cccc|cccc}
w_1& 0& 0& 0& w_2& 0& 0& 0& w_3& 0& 0& 0& w_4& 0& 0& 0 \\
w_{5} & 0 & 0 & 0 & w_{6} & 0 & 0 & 0 & 
w_{7} & 0 & 0 & 0 & w_{8} & 0 & 0 & 0 \\
w_{9} & 0 & 0 & 0 & w_{10} & 0 & 0 & 0 & w_{11} & 0 & 0 & 0 &  
w_{12} & 0 & 0 & 0 \\
w_{13} & 0 & 0 & 0 & w_{14} & 0 & 0 & 0 & 
w_{15} & 0 & 0 & 0 & w_{16} & 0 & 0 & 0 \\ \hline
0 & w_{6} & 0 & 0 & 0 & w_{5} & 0 & 0 & 0 & w_{8} & 0 & 0 & 0 & 
w_{7} & 0 & 0 \\
0 & w_{2} & 0 & 0 & 0 & w_{1} & 0 & 0 & 0 & 
w_{4} & 0 & 0 & 0 & w_{3} & 0 & 0 \\
0 & w_{14} & 0 & 0 & 0 & 
 w_{13} & 0 & 0 & 0 & w_{16} & 0 & 0 & 0 & w_{15} & 0 & 0 \\ 
0 & w_{10} & 0 & 0 & 0 & w_{9} & 0 & 0 & 0 & w_{12} & 0 & 0 & 0 & 
w_{11} & 0 & 0 \\ \hline
0 & 0 & w_{11} & 0 & 0 & 0 & w_{12} & 0 & 0 & 
0& w_{9} & 0 & 0 & 0 & w_{10} & 0 \\
0 & 0 & w_{15} & 0 & 0 & 
0 & w_{16} & 0 & 0 & 0 & w_{13} & 0 & 0 & 0 & w_{14} & 0 \\
0 & 0 & w_{3} & 0 & 0 & 0 & w_{4} & 0 & 0 & 0 & w_{1} & 0 & 0 & 
0& w_{2} & 0 \\
0& 0 & w_{7} & 0 & 0 & 0 & w_{8} & 0 & 0 & 
0 & w_{5} & 0 & 0 & 0 & w_{6} & 0 \\ \hline
0 & 0 & 0 & w_{16} & 0 & 
0 & 0 & w_{15} & 0 & 0 & 0 & w_{14} & 0 & 0 & 0 & w_{13}  \\
0 & 0 & 0 & w_{12} & 0 & 0 & 0 & w_{11} & 0 & 0 & 0 & w_{10} & 0 & 
0 & 0 & w_{9}  \\
0 & 0 & 0 & w_{8} & 0 & 0 & 0 & w_{7} & 0 & 
0 & 0 & w_{6} & 0 & 0 & 0 & w_{5}  \\
0 & 0 & 0 & w_{4} & 0 & 
0 & 0 & w_{3} & 0 & 0 & 0 & w_{2} & 0 & 0 & 0 & w_{1} \\
\end{array}
\right) ,
\end{equation}
}
where the vertex weights $w_i$ are obtained in terms 
of the spin edge weights as follows,
\begin{eqnarray}
\label{WASH}
&& w_1 = 1,~~ 
w_2 = \frac{b_v(x - y)}{b_h(y)},~~ 
w_3=  \frac{c_v(x - y)}{c_h(y)},~~ 
w_4 = \frac{d_v(x - y)}{d_h(y)},~~ 
w_5 = b_h(x) b_v(x - y),~~ \nonumber \\
&& w_6 = \frac{b_h(x)}{b_h(y)},~~ 
w_7 = \frac{b_h(x) d_v(x - y)}{c_h(y)},~~ 
w_8 = \frac{b_h(x) c_v(x - y)}{d_h(y)},~~ 
w_9 = c_h(x) c_v(x - y), \nonumber \\
&& w_{10} = \frac{c_h(x) d_v(x - y)}{b_h(y)},~~ 
w_{11} = \frac{c_h(x)}{c_h(y)},~~ 
w_{12}= \frac{b_v(x - y) c_h(x)}{d_h(y)},~~ 
w_{13} = d_h(x) d_v(x - y), \nonumber \\
&& w_{14}= \frac{c_v(x - y) d_h(x)}{b_h(y)},~~ 
w_{15}= \frac{b_v(x - y) d_h(x)}{c_h(y)},~~ 
w_{16}= \frac{d_h(x)}{d_h(y)} .
\end{eqnarray}

We have verified that the $\mathrm{R}$-matrix (\ref{RASH},\ref{WASH}) 
indeed satisfies 
the Yang-Baxter equation by using the explicit 
expressions of the edge weights (\ref{TETAW}). This
can be done by using certain identities among theta 
functions and with the 
help of symbolic algebra packages.

We would like to conclude this section by discussing 
the Hamiltonian limit  associated to the above
$\mathrm{R}$-matrix when the Ashkin-Teller spin model is layer 
isotropic. In this
case the two independent Ising interactions are the same, i.e $J_h=K_h$ and $J_v=K_v$,
and the Ashkin-Teller model becomes equivalent
to a staggered six-vertex model with $w_d=0$ \cite{BAX,KOMO}. This corresponds
to the limit $q \rightarrow 0$ in Eq.(\ref{TETAW}) and the respective 
weights are
given in terms of trigonometric functions,
\begin{eqnarray}
\label{CRIA}
&& a_h(x)=1,~~b_h(x)=c_h(x)= \frac{\sin \left(\frac{\xi-x}{2} \right)}{\sin \left( \frac{\xi+x}{2}\right)},~~
d_h(x)= \frac{\tan \left(\frac{\xi-x}{2} \right)}{\tan \Big( \frac{\xi+x}{2}\Big)}, \nonumber \\
&& a_v(x)=1,~~b_v(x)=c_v(x)= \frac{\sin \left(\frac{x}{2} \right)}{\sin \left( \xi-\frac{x}{2}\right)},~~
d_v(x)= \frac{\tan \left(\frac{x}{2} \right)}{\tan \left(\xi- \frac{x}{2} \right)} ,
\end{eqnarray}
and from Eq.(\ref{WASH}) we have now 
ten distinct weights $w_i$ due to the identity $b_{h,v}(x)=c_{h,v}(x)$. The respective auxiliary functions
associated to the inversion relations are,
\begin{equation}
\rho_1(x)=1,~~\rho_2(x)= \frac{\sin\left(\xi+x\right) \sin\left(\xi-x \right)}
{\sin\left(\xi+\frac{x}{2}\right) \sin\left(\xi-\frac{x}{2}\right) \Big[\cos\left(\frac{x}{2}\right)\Big]^2} .
\end{equation}

Inspired by the analysis in section 4 we write the Hamiltonian 
associated to the vertex model defined by the weights (\ref{WASH},\ref{CRIA}) as,
\begin{equation}
\label{HAMASH}
H(x_0)= -J\Big[ \sum_{j=1}^{L-1} \left( H^{(0)}_{j,j+1}(x_0) +H^{(1)}_{j,j+1}(x_0) \right) +H^{(0)}_{L,1}(x_0) +H^{(1)}_{L,1}(x_0) \Big] ,
\end{equation}
where the dynamics of the Hamiltonian will be described by the following two commuting 
sets of spin-$\frac{1}{2}$ Pauli matrices, 
\begin{equation}
\sigma^{x}=\left(\begin{array}{cc}  
0 & 1 \\
1 & 0 \\
\end{array}
\right) \otimes \mathrm{I}_2,~~
\sigma^{z}=\left(\begin{array}{cc}  
1 & 0 \\
0 & -1 \\
\end{array}
\right) \otimes \mathrm{I}_2,~~
\tau^{x}=\mathrm{I}_2 \otimes \left(\begin{array}{cc}  
0 & 1 \\
1 & 0 \\
\end{array}
\right),~~
\tau^{z}=\mathrm{I}_2 \otimes \left(\begin{array}{cc}  
1 & 0 \\
0 & -1 \\
\end{array}
\right) .
\end{equation}

We find that the form of the two-body $H^{(0)}_{j,j+1}(x_0)$ term 
is similar to 
that obtained considering a particular time-continuous limit
of the classical Ashkin-Teller model which preserves
the self-dual property of such spin model \cite{KOMO}, namely
\begin{equation}
\label{HAMASH1}
H^{(0)}_{j,j+1}(x_0) = \sigma^z_j \sigma^z_{j+1}+\sigma^x_j+\tau^z_j \tau^z_{j+1} + \tau^x_j+
\frac{\cos(\xi)}{\cos(x_0)} \Big(\sigma^z_j \sigma^z_{j+1} \tau^z_j \tau^z_{j+1} ,
+ \sigma^x_j\tau^x_j \Big)
\end{equation}
while the expression for the additional two-body $H^{(1)}_{j,j+1}(x_0)$ term is given by,
\begin{eqnarray}
\label{HAMASH2}
H^{(1)}_{j,j+1}(x_0)&=&-\frac{\cos(\xi) \sin(x_0)}{\cos(x_0) \sin(\xi)}\Big(\left(\sigma^x_j+\tau^x_j\right) \sigma^z_j \sigma^z_{j+1} \tau^z_{j} \tau^z_{j+1}-\left(\sigma^z_j \sigma^z_{j+1}+\tau^z_{j} \tau^z_{j+1}\right) \sigma^x_j \tau^x_j\Big) \nonumber \\
&+& \frac{\cos(\xi)}{\cos(x_0)} \Big[\frac{\sin(x_0)}{\sin(\xi)}\Big]^2 \Big(\sigma^x_j \tau^z_{j} \tau^z_{j+1}+\tau^x_j \sigma^z_j \sigma^z_{j+1}+\sigma^x_j \tau^x_j \sigma^z_j \sigma^z_{j+1} \tau^z_{j} \tau^z_{j+1} \Big) \nonumber \\
&-& \frac{\sin(x_0)}{\sin(\xi)} \Big(\sigma^x_j \sigma^z_j \sigma^z_{j+1}+\tau^x_j \tau^z_{j} \tau^z_{j+1}\Big) .
\end{eqnarray}

We finally note that the Hamiltonian defined by Eqs.(\ref{HAMASH},\ref{HAMASH1},\ref{HAMASH2}) is a Hermitian operator
when the parameter
$x_0$ is imaginary. 

\section{Conclusions}

The main purpose of this paper was to explore a recent correspondence 
among arbitrary $n$-state spin and vertex models on the square lattice \cite{MAR}
in the realm of exactly solvable two-dimensional systems. We have been
able to formulate a given integrable classical spin model with edge weights
depending on the difference of spectral parameters in the framework 
of an equivalent solvable
vertex model on the square lattice. An immediate consequence of
such embedding is that 
the exact solution of integrable spin models can in principle be considered
within the quantum inverse scattering method \cite{FAD}.

In this sense, we have exhibited the expressions of the Lax operator of the vertex 
model and the 
respective $\mathrm{R}$-matrix in terms of the spin edge weights which are shown to 
satisfy the Yang-Baxter algebra. The unitarity of the $\mathrm{R}$-matrix and therefore
its invertibility is assured by assuming that the spin model edge weights satisfy certain
inversion relation. It turns out that the $\mathrm{R}$-matrix is not of difference form
and this feature can be explored to construct deformed quantum spin chains with additional
interactions generalizing
those associated with the classical integrable spin models. We have applied this 
construction
to the $n$-state scalar Potts \cite{POT} and the Ashkin-Teller \cite{ASH} 
models and the 
expressions of the respective $\mathrm{R}$-matrices 
are presented. In the case of the scalar Potts model
we have argued that the $\mathrm{R}$-matrix can be written in terms of the underlying
Temperley-Lieb operators \cite{TEMP}. We believe that such $\mathrm{R}$-matrices can be
viewed as new solutions of the Yang-Baxter equation without the difference property on
the spectral parameters.

In principle, we can attempt to formulate the $\mathrm{R}$-matrix associated to the  equivalent vertex model
without the need of any
specific assumption about the parametrization of the edge weights of the spin model. To this end let us consider 
two spin models
one of them with edge weights $W_{h,v}(i,j)$ and the other one with distinct prime edge 
weight $W^{'}_{h,v}(i,j)$. The sufficient 
conditions for the commutation
of the diagonal-to-diagonal transfer matrices of these spin models  
consist of the existence of 
double primed weights $W_{h,v}^{''}(i,j)$ which are required to
satisfy the following sets of star-triangle relations,  
\begin{eqnarray}
\label{STARgene}
&& \sum_{d=1}^{n} W^{'}_v(d,c) W^{''}_v(b,d) W_h(a,d)= \mathcal{R} W^{'}_h(a,b) W^{''}_h(a,c) W_v(b,c) , \nonumber \\
&& \sum_{d=1}^{n} W^{'}_v(c,d) W^{''}_v(d,b) W_h(d,a)= \mathcal{R} W^{'}_h(b,a) W^{''}_h(c,a) W_v(c,b) ,
\end{eqnarray}
where again the factor $\mathcal{R}$ is assumed
to be independent of the spin variables
$a,b,c=1,\dots n$. 

The requirement (\ref{STARgene}) leads to $2n^3$ relations which in principle can 
be solved by eliminating 
auxiliary edge weights $W_{h,v}^{''}(i,j)$ and the scalar factor $\mathcal{R}$ and as result
we end up with a number of non-linear equations involving the edge weights $W_{h,v}(i,j)$ and $W^{'}_{h,v}(i,j)$.
We then assume that these non-linear equations can be solved in such a way that both edge weights $W_{h,v}(i,j)$ and $W^{'}_{h,v}(i,j)$ lie on the same algebraic 
variety. After performing
the above tasks the auxiliary weights $W_{h,v}^{''}(i,j)$ and the factor 
$\mathcal{R}$ will be
determined by
polynomials whose variables are the edge weights $W_{h,v}(i,j)$ and $W^{'}_{h,v}(i,j)$. 
Therefore, at this point, we can shorten the notation and represent the set of edge weights  
$W_{h,v}(i,j)$ and $W^{'}_{h,v}(i,j)$ by the symbols 
${\bf w}$ and ${\bf w}^{'}$, respectively. Now we can write the
Yang-Baxter algebra associated with the equivalent vertex model as follows,
\begin{equation}
\label{YBAgen}
\mathrm{R}_{12}({\bf w},{\bf w}^{'}) \mathbb{L}_{13}({\bf{w}}) \mathbb{L}_{23}({\bf{w}}^{'})= 
\mathbb{L}_{23}({\bf{w}}^{'}) \mathbb{L}_{13}({\bf{w}}) \mathrm{R}_{12}({\bf w},{\bf w}^{'}) ,
\end{equation}
where the respective Lax operator is given by Eq.(\ref{LAX}), namely
\begin{equation}
\mathbb{L}_{12}({\bf{w}}) 
=\sum_{i,j,k=1}^{n} W_h(j,i) W_v(j,k) e_{ik} \otimes e_{ji} .
\end{equation}

Considering the reasoning of section 3 but now with the help of 
the star-triangle relations (\ref{STARgene}), it is
possible to show that the following $\mathrm{R}$-matrix,
\begin{equation}
\label{RMAgen}
\mathrm{R}_{12}({\bf w},{\bf w}^{'}) =\sum_{i,j,k=1}^{n} \frac{W_h(j,i) W^{''}_v(j,k|{\bf w},{\bf w}^{'})}{W^{'}_h(k,i)} e_{ik} \otimes e_{ji} ,
\end{equation}
satisfy the Yang-Baxter algebra (\ref{YBAgen}). Note that here we have 
emphasized the  dependence of the auxiliary variables 
$W_{h,v}^{''}(i,j)$ on the spin edge weights $W_{h,v}(i,j)$ and $W^{'}_{h,v}(i,j)$.

The next step for the integrability is to assure that $\mathrm{R}$-matrix has an inverse for most values
of the spin edge weights ${\bf w}$ and ${\bf w}^{'}$. One way to guarantee 
this property is by imposing that
the $\mathrm{R}$-matrix satisfy 
the unitarity property,
\begin{equation} 
\label{UNIgen}
\mathrm{R}_{12}({\bf w},{\bf w}^{'})
\mathrm{R}_{21}({\bf w}^{'},{\bf w})= \rho_2({\bf w},{\bf w}^{'}) \mathrm{I}_n \otimes \mathrm{I}_n ,
\end{equation}
when we interchange the spin model edge weights, that is ${\bf w} \leftrightarrow {\bf w}^{'}$. Considering
the $\mathrm{R}$-matrix expression (\ref{RMAgen}) we find that the unitarity property is satisfied
provided that the vertical
auxiliary weights $W_{v}^{''}(i,j)$ satisfy the following relation,
\begin{equation}
\label{INVgen}
\sum_{k=1}^{n}W^{''}_v(i,k|{\bf w},{\bf w}^{'})
W^{''}_v(k,j|{\bf w}{'},{\bf w}) =\rho_2({\bf w},{\bf w}^{'}) \delta_{i,j}
\end{equation}
which is the analog of the second inversion 
relation given in Eq.(\ref{INV}).

We hope that the above abstract construction could be used to include 
spin models whose edge weights can not be presented in terms of the difference
of two spectral parameters being the most known example the chiral Potts model \cite{MCPERK2}. However,
it could be that such construction still needs further adaptations to include the specific representation of
the edge weights of the chiral Potts model in terms of two distinct points on an algebraic curve, in special attention
to the constraint (\ref{INVgen}).

\section*{Acknowledgments}

This work was supported in part by the Brazilian Research Council CNPq 305617/2021-4.

\addcontentsline{toc}{section}{Appendix A}
\section*{\bf Appendix A: The Yang-Baxter equation}
\setcounter{equation}{0}
\renewcommand{\theequation}{A.\arabic{equation}}

We start by rewriting the Yang-Baxter equation (\ref{YBE}) in terms of its components,
\begin{equation}
\label{YBEE}
\sum_{\gamma,\gamma^{'},\gamma^{''}=1}^{n} \mathrm{R}_{a_1,a_2}^{\gamma,\gamma^{'}}(x,y)
\mathrm{R}_{\gamma,a_3}^{b_1,\gamma^{''}}(x,z)
\mathrm{R}_{\gamma^{'},\gamma^{''}}^{b_2,b_3}(y,z)=
\sum_{\gamma,\gamma^{'},\gamma^{''}=1}^{n} \mathrm{R}_{a_2,a_3}^{\gamma^{'},\gamma^{''}}(y,z)
\mathrm{R}_{a_1,\gamma^{''}}^{\gamma,b_3}(x,z)
\mathrm{R}_{\gamma,\gamma^{'}}^{b_1,b_2}(x,y) .
\end{equation}

By substituting the expression of the $\mathrm{R}$-matrix (\ref{RMA1}) we observe that
Yang-Baxter equation (\ref{YBEE}) is trivially satisfied for $b_3 \neq a_1$ 
since both sides of the equation are in fact zero. The situation is similar to that we have found for the Yang-Baxter algebra, see Eqs.(\ref{LEFT},\ref{RIGHT}). It turns out
that for $b_3=a_1$ the Yang-Baxter equation (\ref{YBEE}) becomes,
\begin{eqnarray}
\label{YBEind}
&& W_v(a_3,b_1|x-z) \sum_{\gamma=1}^{n} \frac{W_v(a_2,\gamma|x-y) W_h(a_3,\gamma|x) W_v(\gamma,b_2|y-z)}{W_h(b_1,\gamma|z)}= \nonumber \\
&& \frac{W_h(a_3,a_2|y)}{W_h(b_1,b_2|y)} W_v(a_2,b_2|x-z) \sum_{\gamma^{'}=1}^{n} \frac{W_v(a_3,\gamma^{'}|y-z) W_h(\gamma^{'},b_2|x) W_v(\gamma^{'},b_1|x-y)}{W_h(\gamma^{'},a_2|z)} ,
\end{eqnarray}
reducing the summations over three different labels into a sum over a single index.

We have checked that the relations (\ref{YBEind}) are satisfied by 
the edge weights of the
scalar Potts, the self-dual Ashkin-Teller, 
the Fateev-Zamolodchikov, 
and the Kashiwara-Miwa spin models.
This has been done by substituting the explicit expressions of the weights 
of the mentioned spin models
in Eq.(\ref{YBEind}). The simplifications can be carried out by considering the addition properties between either trigonometric or elliptic theta functions 
and with the help of symbolic algebra packages. In addition to that, we have used that the edge weights satisfy the property
$W_{h,v}(i,j|x)=W_{h,v}(j,i|x)$ and in this case we note 
that Yang-Baxter relations (\ref{YBEind}) are
trivially satisfied for the subset $b_1=a_2$ and $b_2=a_3$.

We now recall that the weights of the first two mentioned spin models have been already 
discussed in sections 4 and 5. Therefore, for the sake of completeness, we
provide below the edge weights of the Fateev-Zamolodchikov and Kashiwara-Miwa models 
in order to easy independent verifications for
such spin models.

\begin{center}
$\bullet$ The Fateev-Zamolodchikov model
\end{center}

The Fateev-Zamolodchikov model is $n$-state spin model with underlying $Z(n)$ symmetry. The horizontal and vertical weights of this model are,
\begin{equation}
W_h(a,b|x)= \prod_{j=1}^{|a-b|} \frac{\sin \left( (2j-1) \lambda -x \right)}{\sin \left( (2j-1) \lambda +x \right)},~~
W_v(a,b|x)= \prod_{j=1}^{|a-b|} \frac{\sin \left( (2j-2) \lambda +x \right)}{\sin \left( 2j \lambda -x \right)} ,
\end{equation}
where $\lambda=\frac{\pi}{2n}$.

For $n=2,3$ this spin model corresponds to the two and three states scalar Potts model. The normalizations of the inversion relations 
are given by,
\begin{equation}
\rho_1(x)=1,~~\rho_2(x)= n\prod_{j=1}^{[n/2]} 
\frac{\sin \left( (2j-1) \lambda +x \right) \sin \left( (2j-1) \lambda -x \right)  }{\sin \left( 2j \lambda +x \right) \sin \left( 2j \lambda -x \right)}
\end{equation}

\begin{center}
$\bullet$ The Kashiwara-Miwa model
\end{center}

The Kashiwara-Miwa model is a generalization of the Fateev-Zamolodchikov model that 
breaks the $Z(n)$ invariance but retains the weights reflection symmetry $W_{h,v}(i,j)=W_{h,v}(j,i)$ 
as well as the rapidity difference
property. 
This model has also been investigated by Hasegawa and Yamada \cite{HAS} and by Gaudin \cite{GAU}.
The corresponding edge weights are formulated in terms of the theta functions  $\theta_1(x,q)$ and $\theta_4(x,q)$,
\begin{eqnarray}
W_h(a,b|x) &=& [f(a) f(b)]^{-nx/\pi} \prod_{j=1}^{|a-b|} \frac{\theta_1 \left( (2j-1) \lambda -x,q \right)}{\theta_1 \left( (2j-1) \lambda +x,q \right)} 
\prod_{j=1}^{a+b} \frac{\theta_4 \left( (2j-1) \lambda -x,q \right)}{\theta_4 \left( (2j-1) \lambda +x,q \right)} ,  \nonumber \\
W_v(a,b|x) &=& [f(a) f(b)]^{n(x-\lambda)/\pi} \prod_{j=1}^{|a-b|} \frac{\theta_1 \left( (2j-2) \lambda +x,q \right)}{\theta_1 \left( 2j \lambda -x,q \right)} 
\prod_{j=1}^{a+b} \frac{\theta_4 \left( (2j-2) \lambda +x,q \right)}{\theta_4 \left( 2j \lambda -x,q \right)} ,
\end{eqnarray}
where $f(a)=\frac{\theta_4(0,q)}{\theta_4(2\pi a/n,q)}$. 

For $n=2$ this spin model corresponds to the Ising model solved originally by Onsager. The normalizations of the inversion relations 
are given by,
\begin{equation}
\rho_1(x)=1,~~\rho_2(x)=  \frac{h(x)h(-x)}{\left[h(0)\right]^2} ,
\end{equation}
where the function $h(x)$ in terms of the theta functions is,
\begin{equation}
h(x)= \prod_{j=1}^{[n/2]} 
\frac{\theta_1 \left( (2j-1) \lambda +x,q \right) \theta_4 \left( (2j-1) \lambda +x,q \right)  }{\theta_1 \left( 2j \lambda +x,q \right) \theta_4 \left( 2j \lambda +x,q \right)} .
\end{equation}


\begin{thebibliography}{}
%
\bibitem{BAX} R.J. Baxter, {\em Exactly Solved Models in Statistical Mechanics}, Academic Press, 1982
%
\bibitem{ONSA1} L. Onsager, {\em Phys.Rev.}, 65 (1944) 117
%
\bibitem{MCT} B.M. McCoy and T.T. Wu, {\em The Two Dimensional Ising Model}, Havard University Press, 1973
%
\bibitem{KM} H.A. Kramers and G.H. Wannier, {\em Phys.Rev. 60 (1941) 252 }
%
\bibitem{ONSA2} L. Onsager, {\em Critical Phenomena in Alloys, Magnets and Supercondutors}, eds. Mills, Ascher and Jaffee,
McGraw-Hill, New York, 1971
%
\bibitem{MIT} M.J. Stephen and L. Mittag, {\em J.Math.Phys. 13 (1972) 1944}
%
\bibitem{KRA} G.H. Wannier, {\em Rev.Mod.Phys 17 (1945) 50 }
%
\bibitem{PERK} H. Au-Yang and J.H.H Perk {\em Advanced Studies in Pure Mathematics 19, (1989) 57}
%
\bibitem{BAX1} R.J. Baxter, {\em Int.J.Mod.Phys.B 11 (1997) 27}
%
\bibitem{PERK1} J.H.H Perk and H. Au-Yang, {\em Yang-Baxter Equations}, Encyclopedia Vol.5, Elsevier, Amsterdam, 2006
%
\bibitem{POT} R.B. Potts, {\em Proc.Camb.Phil.Soc. 48 (1952) 106}
%
\bibitem{BAX3} R.J. Baxter, {\em J.Phys.C 6 (1973) L445}
%
\bibitem{BASH} V.S. Pokrovsky and Yu.A. Bashilov, {\em Commun.Math.Phys. 84 (1982) 103}
%
\bibitem{ASH} J. Ashkin and E. Teller, {\em Phys.Rev. 64 (1943) 178}
%
\bibitem{WEG} F.J. Wegner, {\em J.Phys.C: Solid State Phys. 5 (1972) L131}
%
\bibitem{FEN} P. Fendley and P. Ginsparg, {\em Nucl.Phys.B 324 (1989) 549}
%
\bibitem{SEA} P. Pearce and K.A. Seaton, {\em J.Phys.A: Math.Gen. 23 (1990) 1191}
%
\bibitem{FAZA} V.A. Fateev and A.B. Zamolodchikov, {\em Phys.Lett.A 92 (1982) 37}
%
\bibitem{KAS} M. Kashiwara and T. Miwa, {\em Nucl.Phys.B 275 (1986) 121}
%
\bibitem{HAS} K. Hasegawa and Y. Yamada, {\em Phys.Lett.A 146 (1990) 387}
%
\bibitem{GAU} M. Gaudin, {\em Journal de Physique I 1 (1991) 351}
%
\bibitem{MAR} M.J. Martins, {\em Nucl.Phys.B 1005 (2024) 116610}
%
\bibitem{BAX2} R.J. Baxter, {\em Ann.Phys. 70 (1972) 193; Ann.Phys. 70 (1972) 323 }
%
\bibitem{FAD} L.A. Takhtadzhan and L.D. Faddev, {\em Russ.Math.Surv. 34 (1979) 11}
%
\bibitem{MCPERK2} R.J. Baxter, J.H.H. Perk and H. Au-Yang, {\em Phys.Lett.A 128 (1988) 138}
%
\bibitem{BASTR} V.V. Bazhanov and Yu.G. Stroganov, {\em J.Stat.Phys. 59 (1990) 799}
%
\bibitem{LEPO} M. de Leeuw and V. Posch, {\em All $4 \times 4$ solutions of the quantum Yang-Baxter equation},
arXiv:2411.18685
%
\bibitem{TEMP} H.N.V. Temperley and E.H. Lieb, {\em Proc.Roy.Soc.A 322 (1971) 251}
%
\bibitem{JON} V.F.R Jones, {\em Int.J.Mod.Phys. B 4 (1990) 701}
%
\bibitem{BERK} E. Berkan, {\em Nucl.Phys.B 215 (1983) 68}
%
\bibitem{MIT1} L. Mittag and M.J. Stephen,  {\em J.Math.Phys. 12 (1971) 441}
%
\bibitem{SOL} J. Solyom and P. Peuty, {\em Phys.Rev.B 24 (1981) 218}
%
\bibitem{VERN} E. Vernier, E. O'Brien and P. Fendley, {\em J.Stat.Mech. (2019) 043107}
%
\bibitem{FAN} C. Fan, {\em Phys.Lett.A 39 (1972) 136}
%
\bibitem{GRA} I.S. Gradshteyn and I.M. Ryzhik, {\em Table of Integrals, Series and Products}, Academic Press, New York, 1980
%
\bibitem{KOMO} M. Kohmoto, M. Den Nijs and L.P. Kadanoff, {\em Phys.Rev.B 24 (1981) 5229}
%
%
%
\end{thebibliography}
\end{document}